\documentclass[aps,prb,
preprint,groupedaddress,showpacs]{revtex4}
\usepackage{amsmath}
\usepackage[T2A]{fontenc}
\usepackage[cp1251]{inputenc}
\usepackage{graphicx}
\usepackage{wrapfig}
\usepackage{amssymb}
\usepackage{bm}

\evensidemargin\oddsidemargin
\begin{document}

\title{Coherent emission of $\gamma $ quanta by synchrotron radiation
excited nuclei: geometry of nearly backward scattering}
\author{G.~V.~Smirnov}
\email{g.smirnov@gmx.net}
\affiliation{National Research Center ``Kurchatov Institute'', 123182 Moscow, Russia}

\begin{abstract}
A possibility of further development of Synchrotron M\"{o}ssbauer
Source (SMS) of $^{57}$Fe 14.4 keV radiation is considered. The
principles and detailed description of the SMS device is given in
Refs. [\onlinecite {Sm2011, JSR}]. The perfect crystal of Iron
Borate, FeBO$_{3}$, is the central element of this device. The
coherent nuclear fluorescence of IB crystal illuminated by
synchrotron radiation produces the sharply directed beam of 14.4
keV M\"{o}ssbauer radiation from the crystal set at the pure
nuclear Bragg reflection. Up to now the low angle scattering
geometry was used for generation of the coherent $\gamma $
radiation. The analysis performed in the present paper shows that
the source of about two times larger power can be obtained when
nearly backward scattering geometry is employed. This result can
be efficiently applied in development of high resolution
spectroscopy using synchrotron radiation.
\end{abstract}

\maketitle

%\section{\protect\bigskip Abstract}

\newpage

\section{Introduction}

\hskip0.7 cm Backward coherent scattering of x-rays and neutrons is a well
known technique of high-resolution x-ray and neutron spectroscopy. General
dynamical effects in resonant backward coherent scattering of M\"{o}ssabuer
radiation were studied in Ref. [\onlinecite {vB1988}]. New interest in
nuclear resonant back-reflection arose in connection with generation of the
coherent $\gamma $ radiation with the help of synchrotron radiation (SR).
The particular case of the coherent emission of $\gamma $ -ray photons by
synchrotron-radiation-excited $^{57}$Fe nuclei in FeB0$_{3}$ crystal was
studied in Refs. [\onlinecite {Sm2011, Pot2012}]. General properties of
nuclear resonant diffraction are well revealed in this case. During free
de-excitation of the nuclei, a nuclear exciton polariton is developing
inside the crystal and generates at the exit of the crystal a coherent $%
\gamma $ -ray beam. The lifetime of nuclear exciton polariton is
of the order or somewhat less than natural lifetime of the nuclear
excited level. In the case of $^{57}$Fe nuclei it is about
$10^{-7}$ s. In the conditions of stationary illumination of the
crystal by SR a permanent intensity beam of $\gamma $ radiation is
generated.

\hskip0.7 cm Up to now the coherent emission of $\gamma $
radiation was studied in the low angle scattering geometry
[\onlinecite {Pot2012,Sm1997,Sm2000,Mitsu2009}]. Considerable interest in
back-reflections arises due to beneficial applications of this
type reflections in $\gamma $-ray optics. There are several
reasons for that. The use of pure nuclear back-reflections with
high reflectivity in an angular range of 100-200 $\mu rad$
expected for back reflections would allow much more M\"{o}ssbauer
radiation to be generated by the synchrotron beam. The necessary
condition for high reflectivity is the fact that the form factor
of the amplitude of nuclear resonant scattering equals unity in
the whole angular range. This is because a nucleus size is much
less than the wave length of $\gamma $ radiation involved. The
reflectivity in back diffraction can be essentially enlarged by
employing so called asymmetric reflections for which the
reflecting planes are inclined with respect to the plane of
crystalline surface. Beside that more freedom is opened to play
with the polarization factor of nuclear amplitude to increase it.

\hskip0.7 cm Strong nuclear reflections exactly in backward direction might
also be applied in the resonator systems of Fabry-Perrot type. Finally, the
practical advantage of the back reflection is the possibility of using much
more simple driving system. To provide energy modulation the crystal should
be mounted on a M\"{o}ssbauer transducer. At the same time diffraction
conditions must be fulfilled. To allow for diffraction in a low angle
scattering geometry, the crystal is to be attached to the moving frame above
the transducer, Fig. 6 in Ref. [\onlinecite {JSR}]. Therefore special care
is needed to balance the frame. When scattering occurs in back direction the
diffracting crystal can be mounted directly on the transducer rod.

\hskip 0.7 cm Based on the above consideration, aim of the present paper is
to investigate the $\gamma $ rays coherent emission function in the geometry
of backward scattering from nuclear array. We first present solutions for
the Fourier component of electric field in the wave scattered by nuclear array in the
vicinity of nuclear resonance and of Bragg angle. After that angular
functions of the coherent emission and spectral composition of the emitted $%
\gamma $ radiation are calculated for a particular case of the nearly
backward reflection.

\section{ Elements of dynamical theory of nuclear resonant diffraction}

\hskip0.7 cm A nuclear array in a crystal represents for M\"{o}ssbauer
radiation a resonating three-dimensional grating, which gives rise to
resonant Bragg diffraction of $\gamma $ -rays. If regularity of the grating
persists over a large volume of the crystal, the multiple scattering of
radiation occurs. Mutual interference of the propagating and Bragg reflected
waves produces a resultant wave field the structure of which is of a
standing-wave type. In this way the standing-wave mode of nuclear exciton
polariton is realized under conditions of Bragg diffraction. At the exit of
the crystal a coherent beam of resonant $\gamma $ radiation is formed. In
general, the Bragg diffraction in large perfect crystals is described by the
dynamical theory accounting for the multiple scattering of radiation by
atoms. A detailed account of the dynamical diffraction theory of nuclear
resonant diffraction is given by Kagan, Trammell, and Hannon [%
\onlinecite{Ka1999, HanTram1999}]. We shortly summarize the theory based on
the solution of the Maxwell equations and apply it to our case. For a space
and time Fourier component of the electric-field vector $\mathbf{E}\left(
\mathbf{k},\omega \right) $, which represents the amplitude of a plane
monochromatic $\gamma $-ray wave having the wave vector $\mathbf{k}$ and
frequency $\omega $, the Maxwell wave equation can be written in the
following form
\begin{equation}
\left[ k^{2}-K^{2}\right] \mathbf{E}\left( \mathbf{k},\omega \right) -%
\mathbf{k}\left( \mathbf{k\cdot E}\left( \mathbf{k},\omega \right) \right) =%
\frac{4\pi i\omega }{c^{2}}\mathbf{j}\left( \mathbf{k},\omega \right) ,
\label{1}
\end{equation}%
where $K=\omega /c$, $c$ is the light velocity in vacuum, $\mathbf{j}\left(
\mathbf{k},\omega \right) $ is the Fourier component of the induced current
density. It has contributions from both the electric and nuclear subsystems.
But our interest is focused on the pure nuclear reflections. In this case
the interference field is created only by nuclear currents and the above
equation is in fact the equation for a space time Fourier component of
nuclear polariton where the radiation field and nuclear excitation are
coupled. The induced nuclear current density represents a quantum mechanical
average over the nuclear ensemble. In the linear in field approximation the
excited nuclear current  is proportional to the electric field 
$\mathbf{j}\left( \mathbf{k},\omega \right) \propto
\eta \cdot \mathbf{E}\left( \mathbf{k},\omega \right) $, where $\eta $ is
the nuclear susceptibility amplitude. Employing this relationship
one can arrive at the Maxwell equation for the field amplitude only. We
shall consider the case of the two-waves diffraction. In this case the two
coherent waves are built up in the crystal, one propagating in the direction
of incident wave, the other in the direction of the diffracted wave. The
Maxwell wave equation splits then into a set of the two equations
\begin{eqnarray}
\left( \frac{k_{0}^{\text{ }2}}{K^{2}}-1\right) E_{0}^{s} &=&\overset{\sim }{%
\eta }_{00}^{ss}E_{0}^{s}+\eta _{01}^{ss^{\prime }}E_{1}^{s^{\prime }}
\notag \\
\left( \frac{k_{1}^{\text{ }2}}{K^{2}}-1\right) E_{1}^{s^{\prime }} &=&\eta
_{10}^{s^{\prime }s}E_{0}^{s}+\overset{\sim }{\eta }_{11}^{s^{\prime
}s^{\prime }}E_{1}^{s^{\prime }}\quad ,  \label{2}
\end{eqnarray}%
where $E_{d}^{s}$ are the scalar electric field amplitudes for a definite
wave polarization $s$ and propagation direction $d$; $s=\sigma ,\pi $ when
the basic polarizations of radiation, $\pi $ and $\sigma $, are involved and
$d=0,1$ for the forward and Bragg scattered waves respectively; $k_{0}^{%
\text{ }}$ and $k_{1}^{\text{ }}$ are the complex wave numbers describing
the coherent waves inside the crystal , $\eta _{dd^{\prime }}^{ss^{\prime
}}\left( \omega \right) $ are the nuclear susceptibility amplitudes
(radiation frequency $\omega =E/\hbar $ , $E$ the incident photon energy),
the amplitudes labeled by tilde include the small additions of the
electronic susceptibility amplitude, $\chi _{dd^{\prime }}$ , $\overset{\sim
}{\eta }_{00}=\eta _{00}+\chi _{00}$ ,\ $\overset{\sim }{\eta }_{11}=\eta
_{11}+\chi _{11}$ , that is actual for the straightforward scattering. \ The
complex wave numbers $k_{0,1}^{\text{ }}$ differ from the absolute value of
the wave vector in vacuum $K$ by only small complex corrections
\begin{eqnarray}
k_{0}^{\text{ }} &=&K\left( 1+\varepsilon _{0}\right)  \notag \\
k_{1}^{\text{ }} &=&K\left( 1+\varepsilon _{1}\right) \quad .  \label{3}
\end{eqnarray}%
In the vicinity of Bragg angle $\varepsilon _{1}=\alpha /2+\varepsilon
_{0}/\beta $, where $\alpha =-2\sin 2\theta _{B}\Delta \theta $ is the
angular parameter proportional to deviation $\Delta \theta $ from Bragg
angle $\theta _{B}$ and $\beta $ is asymmetry parameter, $\beta =\cos \theta
_{0}/\cos \theta _{1}$, where $\theta _{0,1}$ are the angles between the
inward normal to the crystalline entrance surface and the wave vectors $%
\mathbf{k}_{0,1}$ respectively. The value $1/\sin 2\theta _{B}$ is called in
the dynamical theory as Lorentz factor. With the account of the relations
Eq. (\ref{3}) we arrive at the following equations set (neglecting the small
order values)
\begin{eqnarray}
2\varepsilon _{0}E_{0} &=&\overset{\sim }{\eta }_{00}E_{0}+\eta _{01}E_{1}
\notag \\
\left( -2\varepsilon _{0}+\alpha \right) E_{1} &=&\eta _{10}E_{0}+\overset{%
\sim }{\eta }_{11}E_{1}\quad ,  \label{4}
\end{eqnarray}%
where polarization indexes are omitted. The set of homogeneous equations (%
\ref{4}) have a solution for the scalar field amplitudes only if the
determinant formed by their coefficients turns to zero
\begin{equation}
\left[
\begin{array}{c}
\overset{\sim }{\eta }_{00}-2\varepsilon _{0}\text{ \ \ \ \ \ \ \ \ \ \ \ \ }%
\eta _{01} \\
\text{\ \ \ \ \ \ \ \ \ \ }\eta _{10}\text{ \ \ \ }\ \ \ \ \ \ \text{\ }%
\overset{\sim }{\eta }_{11}+2\varepsilon _{0}-\alpha%
\end{array}%
\right] =0\text{ .}  \label{5}
\end{equation}%
Eq. (\ref{5}) determines the dispersion of the electromagnetic waves in the
crystal, giving the complex value $\varepsilon _{0}$ as a function of the
radiation frequency $\omega $ and of the angular deviation $\alpha $ from
the exact Bragg position. There are two roots of the Eq. (\ref{5})%
\begin{equation}
\varepsilon _{0}^{\left( 1,2\right) }=\frac{1}{4}\left\{ \overset{\sim }{%
\eta }_{00}+\beta \overset{\sim }{\eta }_{11}-\alpha \beta \mp \sqrt{\left(
\overset{\sim }{\eta }_{00}-\beta \overset{\sim }{\eta }_{11}+\alpha \beta
\right) ^{2}+4\beta \eta _{01}\eta _{10}}\right\}  \label{6}
\end{equation}%
Correspondingly the set of two equations\ Eq. (\ref{4}) transforms
into a set of four equations. We write down now the set of four
linear equations first for the waves propagating in the primary
direction
\begin{eqnarray}
2\varepsilon ^{\left( 1\right) }E_{0} &=&\overset{\sim }{\eta }%
_{00}E_{0}+\eta _{01}E_{1}  \notag \\
2\varepsilon ^{\left( 2\right) }E_{0} &=&\overset{\sim }{\eta }%
_{00}E_{0}+\eta _{01}E_{1},  \label{7}
\end{eqnarray}%
and then for those propagating in the scattering direction
\begin{eqnarray}
\left( \alpha +2\varepsilon ^{\left( 1\right) }/\beta \right) E_{1} &=&\eta
_{10}E_{0}+\overset{\sim }{\eta _{11}}E_{1}  \notag \\
\left( \alpha +2\varepsilon ^{\left( 2\right) }/\beta \right) E_{1} &=&\eta
_{10}E_{0}+\overset{\sim }{\eta }_{11}E_{1}.  \label{8}
\end{eqnarray}%
Since the equations are homogeneous in order to find solutions for the
scalar amplitudes one has to attract additional relationships between the
fields. These are given by the boundary conditions. The solutions for the
scalar field amplitude can be written in the following way. For the
constituent field, which propagates in the primary direction as
\begin{equation}
E_{0}(t)=E_{0}^{\left( 1\right) }\exp (i\varepsilon ^{\left( 1\right)
}Kt/\cos \theta _{0}+E_{0}^{\left( 2\right) }\exp (i\varepsilon ^{\left(
2\right) }Kt/\cos \theta _{0},  \label{9a}
\end{equation}%
and for that propagating in the direction of the Bragg reflection as%
\begin{equation}
E_{1}(t)=E_{1}^{\left( 1\right) }\exp (i\varepsilon ^{\left( 1\right)
}Kt/\cos \theta _{0}+E_{1}^{\left( 2\right) }\exp (i\varepsilon ^{\left(
2\right) }Kt/\cos \theta _{0},  \label{10a}
\end{equation}%
where the common phase factors are omitted, $t$ is the depth in the crystal,
at the entrance to the crystal $t=0$.

\hskip0.7 cm At the entrance boundary the diffracted wave is not yet built
up, therefore the scalar amplitude of the incident field must be equal to
the sum of scalar amplitudes of the waves propagating in the forward
direction
\begin{equation}
E_{0}^{\left( 1\right) }+E_{0}^{\left( 2\right) }=E_{0}.  \label{9}
\end{equation}%
The boundary condition for the field at the exit surface is based on the
consideration, that the sum of amplitudes\ of the waves propagating in Bragg
direction must be equal zero, because below this boundary there is no
scattering matter
\begin{equation}
E_{1}^{\left( 1\right) }\exp \left( iK\varepsilon ^{\left( 1\right) }T/\cos
\theta _{0}\right) +E_{1}^{\left( 2\right) }\exp \left( iK\varepsilon
^{\left( 2\right) }T/\cos \theta _{0}\right) =0,  \label{10}
\end{equation}%
where $T$ is the thickness of the crystalline platelet, therefore $T/\cos
\theta _{0}$ is the path length for the radiation beam from the entrance to
the exit surfaces, it is here assumed that $\ {Im}[\varepsilon ^{\left(
1,2\right) }]>0.$ Solution of the Eqs. (\ref{7},\ref{8}) for the scalar
amplitudes of the electric field with the account for the boundary
conditions Eqs. (\ref{9},\ref{10}) yields for the refraction index\textsl{\ }$%
\varepsilon ^{\left( 1\right) }$
\begin{eqnarray}
E_{0}^{\left( 1\right) } &=&E_{0}\frac{\left( 2\varepsilon ^{\left( 2\right)
}-\overset{\sim }{\eta }_{00}\right) }{\left( 2\varepsilon ^{\left( 2\right)
}-\overset{\sim }{\eta }_{00}\right) -\left( 2\varepsilon ^{\left( 1\right)
}-\overset{\sim }{\eta }_{00}\right) \exp \left[ -i\left( \varepsilon
^{\left( 2\right) }-\varepsilon ^{\left( 1\right) }\right) KT/\cos \theta
_{0}\right] }  \notag \\
E_{1}^{\left( 1\right) } &=&-E_{0}\frac{\beta \eta _{10}}{2\varepsilon
^{\left( 2\right) }-\overset{\sim }{\eta }_{00}-\left( 2\varepsilon ^{\left(
1\right) }-\overset{\sim }{\eta }_{00}\right) \exp \left[ -i\left(
\varepsilon ^{\left( 2\right) }-\varepsilon ^{\left( 1\right) }\right)
KT/\cos \theta _{0}\right] }  \label{11}
\end{eqnarray}%
and correspondingly for the refraction index $\varepsilon ^{\left( 2\right)
} $
\begin{eqnarray}
E_{0}^{\left( 2\right) } &=&-E_{0}\frac{2\varepsilon ^{\left( 1\right) }-%
\overset{\sim }{\eta }_{00}}{\left( 2\varepsilon ^{\left( 2\right) }-\overset%
{\sim }{\eta }_{00}\right) \exp \left[ i\left( \varepsilon ^{\left( 2\right)
}-\varepsilon ^{\left( 1\right) }\right) KT/\cos \theta _{0}\right] -\left(
2\varepsilon ^{\left( 1\right) }-\overset{\sim }{\eta }_{00}\right) }  \notag
\\
E_{1}^{\left( 2\right) } &=&E_{0}\frac{\beta \eta _{10}}{\left( 2\varepsilon
^{\left( 2\right) }-\overset{\sim }{\eta }_{00}\right) \exp \left[ i\left(
\varepsilon ^{\left( 2\right) }-\varepsilon ^{\left( 1\right) }\right)
KT/\cos \theta _{0}\right] -\left( 2\varepsilon ^{\left( 1\right) }-\overset{%
\sim }{\eta }_{00}\right) }  \label{12}
\end{eqnarray}%
The two pairs of the scalar amplitudes\ $E_{0,1}^{(1)},$ $E_{0,1}^{(2)},$
corresponding to a particular dispersion correction $\varepsilon
_{0}^{\left( 1\right) }$ or $\varepsilon _{0}^{\left( 2\right) },$ are
found. Each field is a function of angular and frequency parameters - $%
E=E\left( \alpha ,\omega \right) $. The amplitude $E_{0}=\sqrt{I_{SR}/\Delta
\omega }$ is in our case the scalar amplitude of the synchrotron radiation
with $I_{SR}$ as the intensity of synchrotron radiation within the frequency
range $\Delta \omega $ selected by the monochromator system. This is a
specific feature of nuclear resonant Bragg reflection that the \textit{%
coherent emission from a crystal containing nuclear array exhibits a
combined angular and energy dependence}.

\hskip 0.7 cm One may consider the two solutions for the interference
wavefield: WF1- $E^{\left( 1\right) }=E_{0}^{\left( 1\right) }+E_{1}^{\left(
1\right) }$ and WF2 - $E^{\left( 2\right) }=E_{0}^{\left( 2\right)
}+E_{1}^{\left( 2\right) },$ presenting standing waves with amplitudes
modulated along the normal to the scattering planes.

\hskip0.7 cm The signs at the square root in the Eq. (\ref{6}) for $%
\varepsilon $ are taken to have $\ {Im}\left[ \varepsilon ^{\left( 2\right)
}-\varepsilon ^{\left( 1\right) }\right] >0.$ Then the exponential factor $%
\left[ -i\left( \varepsilon ^{\left( 2\right) }-\varepsilon ^{\left(
1\right) }\right) KT/\cos \theta _{0}\right] >0$ and in the limit of thick
crystal and small angles of incidence the denominators of the expressions\ in
the Eq. (\ref{11}) for $E_{0,1}^{\left( 1\right) }$ increase infinitely, so
that the contribution from these field components vanishes. On the contrary,
since $\left[ i\left( \varepsilon ^{\left( 2\right) }-\varepsilon ^{\left(
1\right) }\right) KT/\cos \theta _{0}\right] <0,$ the parts with
exponentials in the denominators of $E_{0,1}^{\left( 2\right) }$ disappear
and in the limit we obtain
\begin{eqnarray}
E_{0}^{\left( 2\right) } &=&E_{0},\text{ \ \ \ and \ }  \notag \\
E_{1}^{\left( 2\right) } &=&-E_{0}\frac{\beta \eta _{10}}{\left(
2\varepsilon ^{\left( 1\right) }-\overset{\sim }{\eta }_{00}\right) }.
\label{13}
\end{eqnarray}%
Only one wavefield survives in the semi-infinite crystal. In the following
section we shall apply the general solutions given by the Eqs. (11,12) to
calculate the angular and spectral properties of $\gamma $ radiation emitted
by the synchrotron radiation excited nuclei in FeB0$_{3}$ crystal in the
geometry of a nearly backward scattering.

\section{ Backward scattering geometry}

\hskip0.7 cm Now we shall try to formulate conditions for reaching as high
as possible coherent response of nuclear array\ excited by synchrotron
radiation. We shall rely on Eq. (\ref{13}) to find the amplitude of the
reflected radiation near the Bragg angle in the nuclear resonance range.
When the crystal is set at the angle $\alpha =\chi _{00}/\beta +\chi _{11}$
the expression for $\varepsilon ^{\left( 1,2\right) }$ Eq. (\ref{6}) takes
the form%
\begin{equation}
\varepsilon _{0}^{\left( 1,2\right) }=\frac{1}{4}\left\{ \eta _{00}+\beta
\eta _{11}\mp \sqrt{\left( \eta _{00}-\beta \eta _{11}\right) ^{2}+4\beta
\eta _{01}\eta _{10}}\right\} .  \label{14}
\end{equation}%
To get the largest strength of nuclear system response one should select
those reflections for which the suppression of the incoherent channels, i.e.
the cancellation of the total amplitude for the individual nuclear
excitation (the Kagan-Afanas'ev rule) is satisfied, see Refs. [%
\onlinecite{Ka1999,Sm1988}]. When the Kagan-Afanas'ev rule is fulfilled, the
following relation between the matrix elements of nuclear susceptibility is
valid $\eta _{00}\eta _{11}=\eta _{01}\eta _{10}$. If this necessary
condition is met $\varepsilon ^{\left( 1\right) }$ turns to zero and the
expression for the field amplitude reduces to%
\begin{equation}
E_{1}^{\left( 2\right) }=E_{0}\frac{\beta \eta _{10}}{\overset{\sim }{\eta }%
_{00}}.  \label{15}
\end{equation}%
Making use this expression one can characterize the properties of strong
reflection. First of all one can see that the field amplitude is
proportional to the asymmetry factor $\beta $. In the symmetric Bragg
geometry $\beta =-1$ (for definition of $\beta $ see text after Eq. (\ref{3}%
)). Obviously, in the conditions of a low angle diffraction this factor can
not be essentially enlarged, while the geometry of back diffraction can
suggest such a possibility. The next factor of proportionality is the
nuclear susceptibility $\eta _{10}$. As it was shown in Ref. [\onlinecite
{Sm1980}] the nuclear susceptibility in the case of pure nuclear diffraction
in FeBO$_{3}$ is proportional to $\cos \left( \theta _{B}-\varphi \right) $
or $\cos \left( \theta _{B}+\varphi \right) $ depending on what polarization
has the incident radiation $\sigma $ or $\pi $, where $\varphi $ is
inclination angle of the reflecting planes to the crystalline surface. The
FeBO$_{3}$ crystals, as grown, usually have the form of plane parallel
platelets with the entrance and exit surfaces parallel to the crystalline
planes (1\:1\:1). For such crystals the best candidate for strong nuclear
resonant scattering of synchrotron radiation is nearly back reflection
(3\:3\:11). It is characterized by Bragg angle $\theta _{B}=84.6^{0}$, i.e., the
scattering angle in this case is $169.2^{0}$. This reflection was studied
earlier in Ref. [\onlinecite{vB1988}]. The chosen crystalline planes (3\:3\:11)
are inclined at an angle of $59.6^{0}$ to the surface of the crystalline
platelet. The scattering scheme is shown in Fig. 1.
\begin{figure}[tph]
{\large \centering  \vskip -2.1cm \includegraphics[width=1.01%
\textwidth]{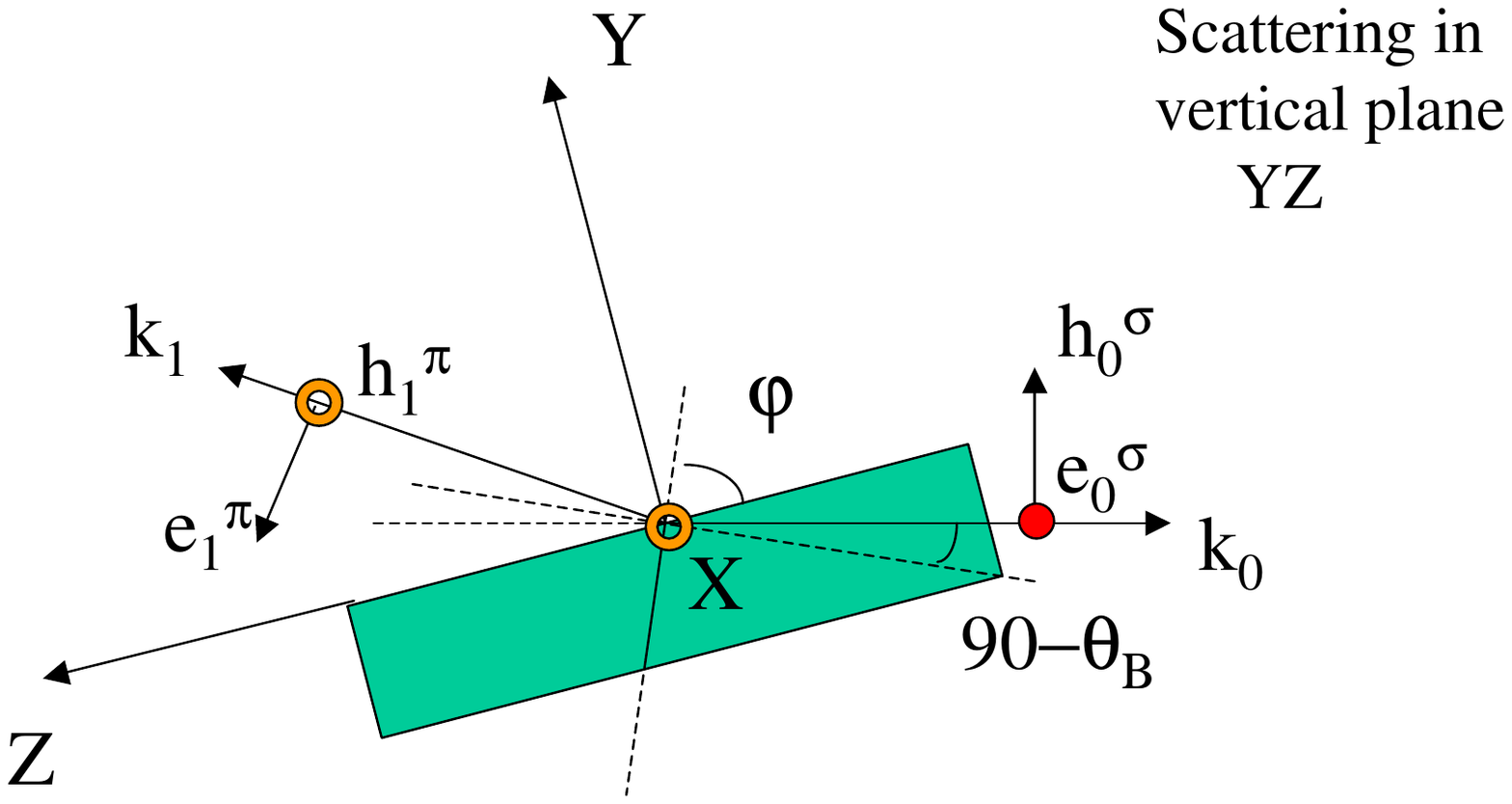} \vskip -4.1cm }
\caption{The scattering geometry for (3\:3\:11) Bragg reflection of 14.4 keV SR
from Iron Borate. The crystal scattering planes are perpendicular to the
vertical plane - the plane of figure, $\protect\varphi =59.6^{0}$. The wave
vectors of incident and scattered radiation radiation $\mathbf{k_{0}}$ and $%
\mathbf{k_{1}}$ form the plane of scattering. This is the plane of figure.
The incident radiation wave vector $\mathbf{k_{0}}$ lies in horizontal
plane. Its magnetic polarization vector $\mathbf{h_{0}^{\protect\sigma }}$
oscillates in the plane of figure along the vertical axis. Since it lies in
the plane of scattering, the incident radiation should be considered as $%
\protect\sigma $-polarized. Due to antiferromagnetic nature of pure nuclear
reflection in Iron Borate the reflected beam is $\protect\pi $-polarized.
Its magnetic vector $\mathbf{h_{1}^{\protect\pi }}$ oscillates
perpendicularly to the scattering plane. }
\label{Fig. 1}
\end{figure}
The asymmetry factor ${\large \beta }$ is 1.369 in the selected scattering
geometry. The scattering in the vertical plane is favorable by several
reasons among which the most important is the possibility to provide the
maximum amplitude of nuclear excitation due to the most conducive value of
the polarization factor of the scattering amplitude. In the case of the
reflection (3\:3\:11) $\cos \left( \theta_{B}-\varphi \right) =0.91$. The
Lorentz factor for this reflection, - 5.34, provides quite large width of
the angular range with the high reflectivity.

\hskip 0.7 cm For calculation of the emission angular dependence and energy
spectra of $\gamma $ rays the codes were made up where the thickness of the
crystal was taken into account in accord with the Eqs. (\ref{11},\ref{12}).
The next figure shows angular emission function for the crystals of various
thicknesses. The low angle reflections (1\:1\:1), Bragg angle $5.1^0$, and
(3\:3\:3), Bragg angle $15.4^0$, are compared with the nearly backward
reflection (3\:3\:11), $\theta_{B}=84.6^0$.

\subsection{ Angular properties, thickness dependence}

\hskip0.7 cm It is supposed that angular distribution of the incident
radiation beam obeys the Gauss law with a characteristic width of 5 $\mu $%
rad that is much less than the angular emission range. The crystal is set at
different angular positions with respect to the center of distribution of
incident beam near the Bragg angles for the reflections (1\:1\:1), (3\:3\:3) and
(3\:3\:11) respectively. The curves depicted in Fig. 2 represent the angular
variation of the emission intensity over the range of scan.
\begin{figure}[tph]
{\large \centering  \vskip -1.1cm \includegraphics[width=0.7%
\textwidth]{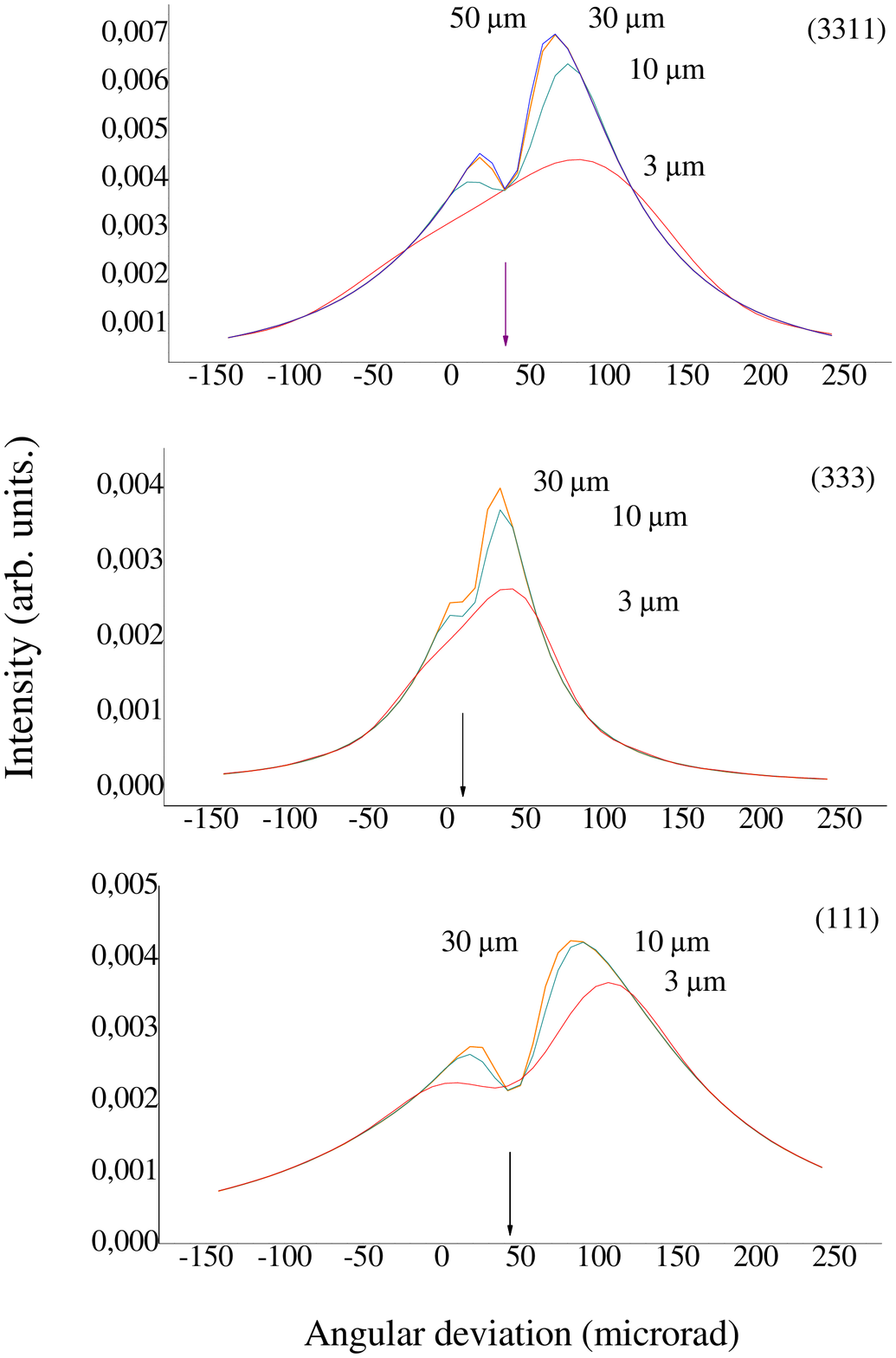} \vskip -1.1cm }
\caption{The angular distributions of $\protect\gamma $ rays emitted by the
IB crystals of various thicknesses heated up to T-Ne\'{e}l (the internal field 2
kOe) for reflections (1\:1\:1), (3\:3\:3) and (3\:3\:11). }
\label{Fig. 2}
\end{figure}
The temperature of crystal is elevated up to Ne\'{e}l temperature where
combined action of the dipole magnetic and the quadrupole electric hyperfine
interaction produces pseudo single line M\"{o}ssbauer diffraction spectrum [%
\onlinecite {Sm2000}]. Integration over the whole resonance range is
fulfilled. The function presenting the emission angular dependence is given
by the following expression
\begin{equation}
I\left( \Delta \theta \right) =\int\limits_{\phi }\int\limits_{\omega }d\phi
d\omega F\left( \phi \right) I\left( \Delta \theta +\phi ,\omega \right)
\text{ },  \label{16}
\end{equation}%
where
\begin{equation}
{\large I}\left( \Delta \theta ,\omega \right) {\large =}\left\vert
E_{1}^{\left( 1\right) }\left( \Delta \theta ,\omega \right) +E_{1}^{\left(
2\right) }\left( \Delta \theta ,\omega \right) \right\vert ^{2},  \label{17}
\end{equation}

\begin{equation}
{\large F}\left( \phi \right) {\large =}\frac{1}{\sigma \sqrt{2\pi }}\exp
\left\{ -\frac{\phi ^{2}}{2\sigma ^{2}}\right\} .  \label{18}
\end{equation}%
For definition of $E_{1}^{\left( 1\right) }$ and $E_{1}^{\left(
2\right) }$ see Eqs. (\ref{11}) and (\ref{12}), for definition of
$\omega $ and $\Delta \theta $ see the text between Eqs. (\ref{2})
and (\ref{3}) and immediately after the Eq. (\ref{3}).

\hskip 0.7 cm As it was found earlier [\onlinecite{Sm2011}], the
multispace quantum interference (involving geometrical energy and
spin domains) in the case of combined magnetic dipole-electric
quadrupole interaction of nuclear spin with crystalline fields in
Iron Borate yields a non-trivial double-hump shape of emission
angular curve. However, Fig. 2 shows, that such a shape is formed
only in the limit of semi-infinite thickness of the crystal. The
path length of radiation in the crystal is determined by
expression $T/\cos \theta _{0}$, where $T$ is the crystal
thickness and $\theta _{0}$ is the angle between the wave vector
$\mathbf{k}_{0}$ and inward normal to the crystalline entrance
surface. Angle $\theta_{0}$ takes the meanings 84.9, 74.5 and 65
degrees for reflections (1\:1\:1), (3\:3\:3) and (3\:3\:11) respectively. So
that the limit of semi-infinite crystal should be reached earlier
for the reflection (1\:1\:1). Indeed, already at the thickness of crystal of
about 3 $\mu $m there are signs of the double-hump structure while
at 10 $\mu $m the saturation in thickness is reached for this reflection 
(see the bottom panel in Fig. 2). In the cases
of both (3\:3\:3) and (3\:3\:11) reflections the emission angular curve
for the thickness of 3 $\mu $m looks as an asymmetric single line.
For these reflections the saturation is reached between 10 and 30
$\mu $m and at 30 $\mu $m respectively. The dip positions are
given by the expression $\Delta \theta =\frac{\chi _{00}/\beta
+\chi _{11}}{2\sin 2\theta _{B}}$ (see the text after Eq. (\ref{3})
and before Eq. (\ref{14})). These are angular positions of the
crystal for Bragg reflections where corrections for the refraction
of radiation wave at the entrance to the crystal due to electronic
scattering are taken into account. For the reflections (1\:1\:1),
(3\:3\:3) and (3\:3\:11) $\Delta \theta $ equals 45, 15, 35 $\mu $rad
respectively. The widths of the curves are largest and about the
same value for reflections (1\:1\:1) and (3\:3\:11) where Lorentz factors
are 5.62 and 5.34 respectively, but the reflection strength in the
case of reflection (3\:3\:11) is much higher. We turn to this question
in the next paragraph.

\subsection{\protect\large Spectral properties of $\protect\gamma $ radiation%
}

\hskip0.7 cm On Fig. 3 the spectra of $\gamma $ radiation emitted by the
crystal are shown for the reflections considered. Integration over angle
through the whole angular range of emission is performed (see Fig. 2)
\begin{equation}
I(E)=\int\limits_{\theta _{1}}^{\theta _{2}}d\theta I\left( \theta ,E\right)
,  \label{19}
\end{equation}%
{with $\theta _{1}=-150\mu $rad and $\theta _{2}=250\mu $rad. }
\begin{figure}[tph]
{\large \centering  \vskip -0.1cm \includegraphics[width=0.75%
\textwidth]{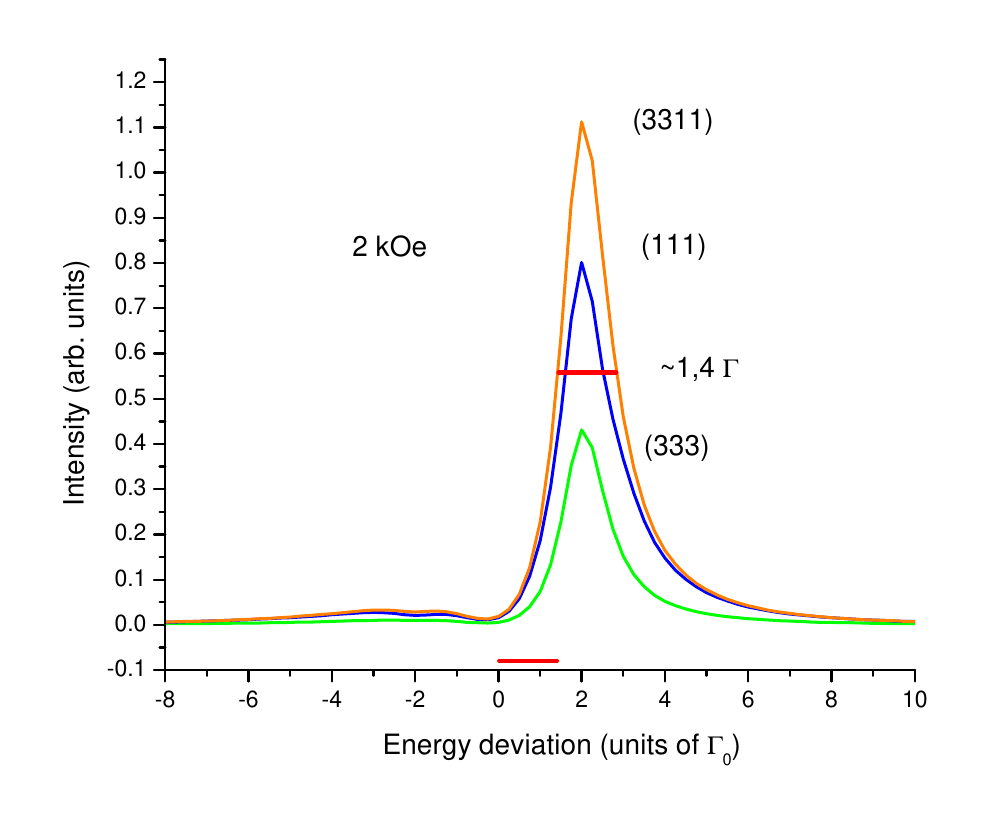} \vskip -0.8cm }
\caption{The energy spectra of $\protect\gamma $ rays emitted by the IB
crystal heated up to T-Ne\'{e}l (the internal field 2 kOe) for reflections
(1\:1\:1), (3\:3\:3) and (3\:3\:11). Nuclear excitation occurs through the whole angular
region of dynamical diffraction. }
\label{Fig. 3}
\end{figure}
The spectra integrated over the broad angular range have the form of
slightly asymmetric single lines with the half widths of about $1.5\Gamma
_{0}$, where $\Gamma _{0}$ is the natural width of nuclear level in the
excited state. As seen from the figure, the emission intensity is the
highest for (3\:3\:11) reflection. It exceeds the emission intensity in the
(1\:1\:1) and (3\:3\:3) reflections by factor of 1.4 and 2.8 respectively. Thus the
back reflection turns to be much more powerful than the reflections at low
angles. The amplitudes of the coherent scattering for (3\:3\:11) reflection in
Iron Borate are presented in Table 1. One can see that the main
contribution to the emission line on the top panel of Fig. 3 is given by 3d
and 6th nuclear transitions between ground and the excited states (the most
intensive transitions and the relevant matrix elements of nuclear
susceptibility are marked in bold).
\begin{table}[tbp]
\caption{Complex amplitudes for nuclear resonant reflection (3\:3\:11) from FeBO$_{3}$ crystal heated up to T-Ne\'{e}l (the internal field 2 kOe) for six nuclear transitions between ground and the first excited states of $^{57}$Fe. }
\label{tab1}%
\begin{ruledtabular}
\begin{tabular}{c|c|c|c|c|c|c|c}
Resonance lines & 1 & 2 & $\mathbf{3}$ & 4 & 5 & $\mathbf{6}$ \\
\hline
Relative positions of \\ lines in units of $\Gamma_0$ & -2.0843 & -2.0841 &
$\mathbf{1.7011}$ & -1.8407 & 1.9445 & $\mathbf{2.2239}$  \\
\hline
Complex amplitudes  & Re \hskip 0.5 cm Im & Re \hskip 0.5 cm Im & Re \hskip 0.5 cm Im & Re \hskip 0.5 cm Im &
Re \hskip 0.5 cm Im & Re \hskip 0.5 cm Im \\
\hline
$\eta _{00}^{\sigma \sigma}$ &{\footnotesize  -19.91  \hskip 0.1 cm  0.00 } & {\footnotesize -4.40 \hskip 0.1 cm  0.00}
&{\footnotesize  $\mathbf{-25.60}$ \hskip 0.1 cm  0 .00 } & {\footnotesize  -18.54 \hskip 0.1 cm 0.00 }
& {\footnotesize -1.58 \hskip 0.1 cm 0.00} & {\footnotesize  $\mathbf{-25.60}$ \hskip 0.1 cm 0 .00 } \\
\hline
$\eta _{01}^{\sigma \pi}$ &{\footnotesize  0.00  \hskip 0.1 cm  -0.80 } & {\footnotesize -0.00 \hskip 0.1 cm  0.00}
&{\footnotesize  0.00 \hskip 0.1 cm  $\mathbf{-13.36}$ } & {\footnotesize  0.00 \hskip 0.1 cm -0.72 }
& {\footnotesize 0.00 \hskip 0.1 cm 0.00}   &   {\footnotesize  0.00 \hskip 0.1 cm $\mathbf{+14.88}$ } \\
\hline
$\eta _{10}^{\pi \sigma}$ &{\footnotesize  0.00  \hskip 0.1 cm  +0.80 } & {\footnotesize -0.00 \hskip 0.1 cm  0.00}
&{\footnotesize  0.00 \hskip 0.1 cm  $\mathbf{+13.36}$ } & {\footnotesize  0.00 \hskip 0.1 cm +0.72 }
& {\footnotesize 0.00 \hskip 0.1 cm 0.00}   &   {\footnotesize  0.00 \hskip 0.1 cm $\mathbf{-14.88}$ } \\
\hline
$\eta _{11}^{\pi \pi}$ &{\footnotesize  -0.03  \hskip 0.1 cm  0.00 } & {\footnotesize 0.00 \hskip 0.1 cm  0.00}
&{\footnotesize  $\mathbf{-6.97}$ \hskip 0.1 cm  0 .00 } & {\footnotesize  -0.03 \hskip 0.1 cm 0.00 }
& {\footnotesize 0.00 \hskip 0.1 cm 0.00} & {\footnotesize  $\mathbf{-8.65}$ \hskip 0.1 cm 0 .00 } \\
\hline
\end{tabular}
\end{ruledtabular}
\end{table}
One can easily check that the Kagan-Afanas'ev rule is fulfilled for these
nuclear transitions. It is also seen, that the resonant energies for these
transitions are nearly coincide, difference between them is only ~ 0.5$%
\Gamma_0 $. Therefore the interference of the considered nuclear transitions
provides a single line emission spectrum, find details in Ref. [\onlinecite
{Sm2011}]. The same physical reason provides the single line emission
spectra in (1\:1\:1) and (3\:3\:3) reflections. {\large {\
\begin{figure}[tph]
{\large \centering  \vskip -0.7cm \hskip -2.35cm \includegraphics[width=0.7%
\textwidth]{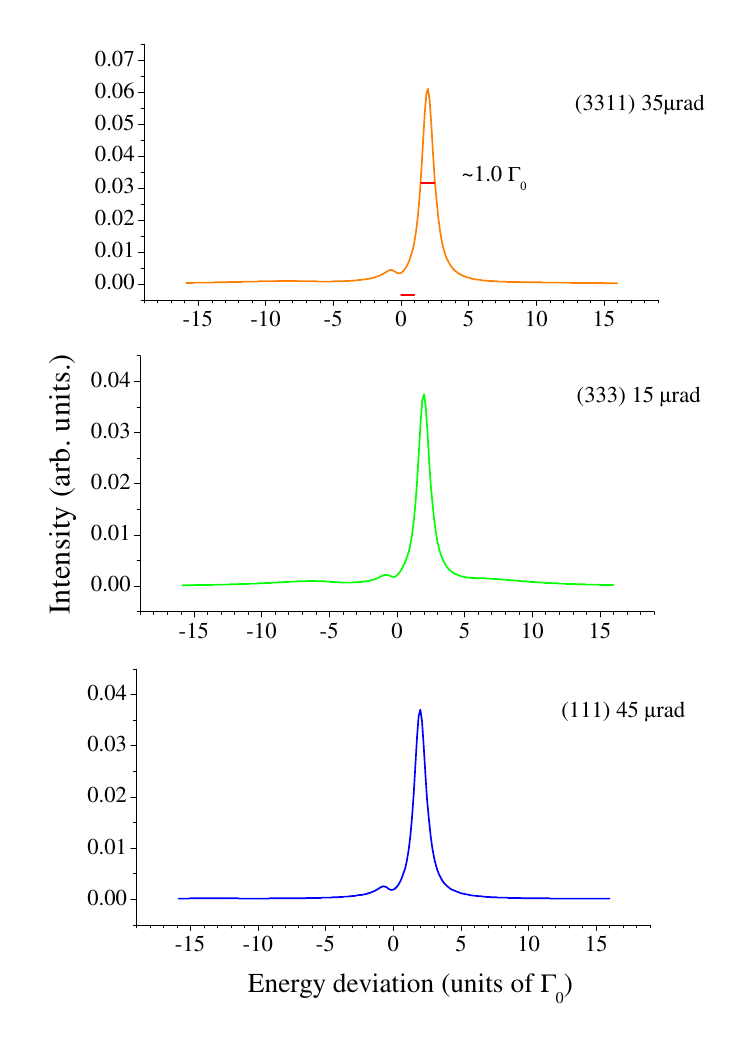} \vskip -1.0cm  }
\caption{The energy spectra of $\protect\gamma $ rays emitted by the IB
crystal heated up to T-Ne\'{e}l (the internal field 2 kOe) for reflections
(1\:1\:1), (3\:3\:3) and (3\:3\:11). Nuclear excitation occurs within angular interval
of 5 $\protect\mu $rad when crystal is set at Bragg angle (dip positions in
Fig. 2 ). }
\label{Fig. 4}
\end{figure}
}}

{\large {\
\begin{figure}[tph]
{\large \centering \vskip -0.7cm \hskip -2.35cm \includegraphics[width=0.7%
\textwidth]{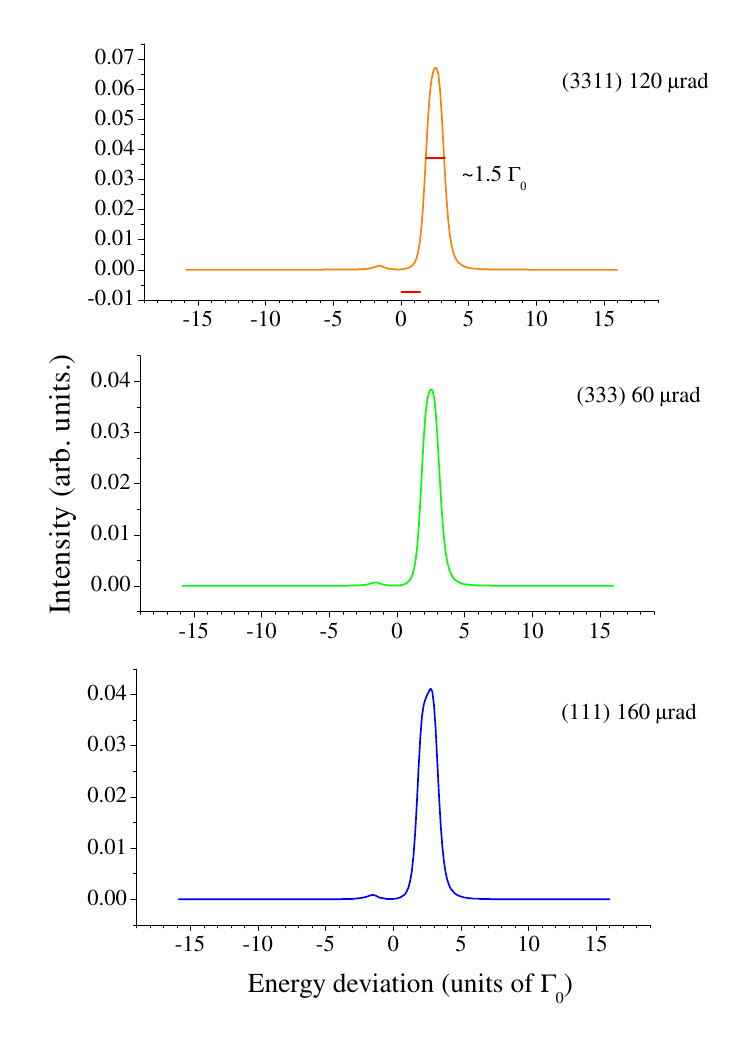} \vskip -1.1cm  }
\caption{The energy spectra of $\protect\gamma $ rays emitted by the IB
crystal heated up to T-Ne\'{e}l (the internal field 2 kOe) for reflections
(1\:1\:1), (3\:3\:3) and (3\:3\:11). Nuclear excitation occurs within angular interval
of 5 $\protect\mu $rad when crystal is set at the middle of right-hand slope
of the angular dependence (see Fig. 2).}
\label{Fig. 5}
\end{figure}
}} \vskip-1.0 cm \hskip0.7cm In the real conditions the exciting synchrotron
radiation is strongly collimated. The divergence of incident beam is usually
much less than the angular range of coherent emission, - somewhat about 5-10
$\mu $rad against 100 $\mu $rad, when the high resolution monochromator is
used. At a fixed angle of incidence of SR the spectrum of the emitted $%
\gamma $ radiation is formed by integration within the narrow angular
interval. In Figs. 4 and 5 the emission spectra are displayed where
convolution of the angular distribution of the incident radiation with the
emission angular distribution is performed for particular settings of the
crystal $\Delta \theta _{p}$ within the emission angular range, as described
by the following formulas%
\begin{equation}
I_{p}\left( E\right) =\int\limits_{\phi }d\phi F\left( \phi \right) I\left(
\Delta \theta _{p}+\phi ,E\right) \text{ ,}  \label{20}
\end{equation}%
\begin{equation}
{\ F}\left( \phi \right) {\large =}\frac{1}{\sigma \sqrt{2\pi }}\exp \left\{
-\frac{\left( \Delta \theta _{p}+\phi \right) ^{2}}{2\sigma ^{2}}\right\} .
\label{21}
\end{equation}%
Fig. 4 shows the spectra of the emitted $\gamma $ radiation in the
positions of the crystal where the dip of the emission angular
function is situated, i. e., exactly in the Bragg positions. As it
was said above, these positions are 15, 35 and 45 $\mu $rad for
reflections (3\:3\:3), (3\:3\:11) and (1\:1\:1) respectively. For the beam of
the exciting radiation strongly limited near these angles there is
an advantage to get a sharper the $\gamma $ - rays source line.
When the crystal is set in the angular position of the dip the
half width of the line is only 1$\Gamma _{0}$. This will allow to
reach
noticeably higher resolution in investigations of hyperfine structure of M%
\"{o}ssbauer spectra. The advantage in the intensity of $\gamma $
radiation generated in the case of (3\:3\:11) reflection evident, the
emitted intensity is by about 1.8 times higher the intensity in
the cases of (1\:1\:1) and (3\:3\:3) reflections. This result is in a good
agreement with the theoretically
estimated benefit at Bragg angle: $\left\{ \left\vert \beta \eta _{10}/%
\overset{\sim }{\eta }_{00}\right\vert ^{2}\right\} _{\left( 3\:3\:11\right)
}/\left\{ \left\vert \beta \eta _{10}/\overset{\sim }{\eta }_{00}\right\vert
^{2}\right\} _{\left( 1\:1\:1\right) }$ $\approx 2$, where the Eq. (\ref{15})
and relevant data for the susceptibilities and the asymmetry factors are
used.

\hskip 0.7 cm If the sharpness of the line source is not that
critical another angular setting of the crystall also provides a
single line spectrum. This position is located in the middle of
the slope on the right-hand side of angular dependence on Fig. 3.
The energy distributions corresponding to this angular setting for
different reflections are shown in Fig. 5. The nature of the
emission line in this angular position is more complicated. A
three-dimensional landscape of a Bragg reflection in the vicinity
of the nuclear resonance and of the Bragg angle near Ne\'{e}l
temperature is presented in Fig. 4c in Ref. [\onlinecite
{Sm2011}]. The two main contributions to the interference pattern
in the range of positive angular deviations from Bragg angle
(corresponding to the right-hand slope in Fig. 2 of the present
paper) come from the long lasting wing of the 6th resonance line
(provided by the real part of the scattering amplitude) and the
line in the center of the resonance range. On the slope of the
angular distribution they together form a slightly broadened and
asymmetric emission line. The evolution of the emission spectrum
with the change of angular
position was investigated experimentally in Ref. [\onlinecite {Pot2012}]
in the case of (3\:3\:3) reflection  and shown in Fig. 3 overthere. 
The emission intensity gain in the case of backward
geometry is preserved at different settings of the crystal in the emission angular range.

%\section{Conclusion}

\hskip 0.7 cm In summary, a possibility of further development of
Synchrotron M\"{o}ssbauer Source (SMS) of $^{57}$Fe 14.4 keV
radiation was found. The formfactor equal unity in the nuclear
resonance scattering for all scattering angles, the possibility to
enlarge the asymmetry and the polarization factors as well as the
Lorentz factor make the back pure nuclear reflections useful for
generation of nuclear resonant $\gamma $ radiation by pure nuclear
coherent scattering of synchrotron radiation. Particularly, in the
case of the nearly back reflection (3\:3\:11) from FeBO$_3$
crystalline platelet having the planes (1\:1\:1) parallel to the
surface the gain in the intensity of generated M\"{o}ssbauer
radiation can reach factor of two compared to the strong low angle
reflection (1\:1\:1) from the same crystal.

The essential practical advantage is that in backscattering geometry the M\"{o}ssbauer
drive system constructed for the case of low-angle scattering [\onlinecite {JSR}] can be
simplified, and namely, be made similar to that used in a regular M\"{o}ssbauer
spectrometer.

\hskip 0.7 cm The author is deeply grateful to Dr. A. I. Chumakov for friendly
motivation and fruitful discussions of this work.

\end{document}